\documentclass[useAMS,usenatbib]{mn2e} 
\usepackage{aas_macros}
\usepackage{graphics}
\usepackage[pdftex]{graphicx}
\usepackage{epstopdf}
\usepackage{epsfig}  
\usepackage{natbib} 
\usepackage{float}
\usepackage{amsmath}
\usepackage{times}
\usepackage[varg]{txfonts}
\usepackage{verbatim} 
\usepackage{multirow,bigdelim} 
\usepackage{color}
\usepackage{vmargin}

\setmarginsrb{1.2cm}{1cm}{1.2cm}{1cm}{1cm}{1cm}{1cm}{1cm}

\title[Bimodality]{Bimodality of Galaxy Disk Central Surface Brightness Distribution in the Spitzer 3.6 $\mu$m band}

\author[Sorce et al.]
{
Jenny G. Sorce$^{1,2}$\thanks{E-mail: \texttt{j.sorce@ipnl.in2p3.fr}}, 
H\'el\`ene M. Courtois$^{1}$,
Kartik Sheth$^{3}$,
R. Brent Tully$^{4}$\\
$^{1}$Universit\'e Lyon 1, CNRS/IN2P3, Institut de Physique Nucl\'eaire, Lyon, France\\
$^2$Leibniz-Institut f\"{u}r Astrophysik, Potsdam, Germany\\
$^3$National Radio Astronomy Observatory/NAASC, 520 Edgemont Road, Charlottesville, VA 22903, USA\\
$^4$Institute for Astronomy, University of Hawaii, 2680 Woodlawn Drive, HI 96822, USA\\
}

\begin{document}

\date{}

\pagerange{\pageref{firstpage}--\pageref{lastpage}} \pubyear{2012}

\maketitle

\label{firstpage}

\begin{abstract}
\indent We report on measurements of the disk central surface brightnesses ($\mu_0$) at 3.6 microns for 438 galaxies selected by distance and absolute magnitude cutoffs from the 2350+ galaxies in the \textit{Spitzer Survey of Stellar Structure in Galaxies} ($S^4G$), one of the largest and   deepest homogeneous mid-infrared datasets of nearby galaxies. Our sample contains nearly 3 times more galaxies than the most recent study of the $\mu_0$ distribution. We demonstrate that there is a bimodality in the distribution of $\mu_0$. Between the low and high surface brightness galaxy regimes there is a lack of intermediate surface brightness galaxies.\\
\indent
Caveats invoked in the literature from small number statistics to the knowledge of the environmental influences, and possible biases from low signal to noise data or corrections for galaxy inclination are investigated.  Analyses show that the bimodal distribution of $\mu_0$ cannot be due to any of these biases or statistical fluctuations. It is highly probable that galaxies settle in two stable modes: a dark matter dominated mode where the dark matter dominates at all radii - this gives birth to low surface brightness galaxies - and a baryonic matter dominated mode where the baryons dominate the dark matter in the central parts - this gives rise to the high surface brightness disks.  The lack of intermediate surface brightness objects suggests that galaxies avoid (staying in) a mode where dark matter and baryons are co-dominant in the central parts of galaxies.
\end{abstract}



\section{Introduction}

Observational surveys of the distribution of galaxy parameters in diverse environments provide essential constraints for theoretical models of galaxy formation and evolution \citep[e.g.][]{2003Ap&SS.284..353T}. The sample selection in these  surveys needs to be as bias-free as possible to conduct an appropriate analysis. Recently a number of large surveys \citep[e.g.][]{2002AJ....123..485S,2009ApJ...703..517D,2010PASP..122.1397S} have been conducted with various telescopes that have led to multiple findings of bimodal distributions of galaxies in terms of color \citep{2004ApJ...600..681B,2006MNRAS.372..253M,2009ApJ...706L.173B,2011ApJ...735...86W}, star formation \citep{2012MNRAS.424..232W} and disk central surface brightnesses \citep[$\mu_0$, e.g.][]{2009MNRAS.393..628M,2009MNRAS.394.2022M}. \citet{2008MNRAS.385.1835B} have even suggested a trimodal distribution for galaxy concentrations.\\

Surface brightness profiles were first studied in 1948 by De Vaucouleurs and later on by \citet{1959Obs....79...54S} and \citet{1970ApJ...160..811F}.  However, the first convincing evidence of a $\mu_0$ bimodal distribution was published only a long time after by \citet{1997ApJ...484..145T}.  This study revealed that the $\mu_0$ distribution in Ursa Major was discontinuous with a lack of galaxies of intermediate surface brightness, or alternatively that there was an excess of low and high surface brightness (L/HSB) galaxies. The authors suggested two stable modes for galaxy formation: A dominant dark matter component at all radii giving birth to LSB galaxies and a dominant baryonic matter in the center giving rise to HSB galaxies. They suggested that the low number of ISB galaxies could be the result of galaxies avoiding the situation where baryonic and dark matters are co-dominant in the center. The very few ISB galaxies could also indicate that there is a small probability for LSB galaxies to turn into HSB galaxies at some point.  However, the authors expressed concerns because of possible large errors in fitting galaxy disks due to shallow K'-band observations which could lead to premature truncation of disks. As a result bulges could be partially included in fits leading to a bias in $\mu_0$. Later \citet{2000MNRAS.311..668B} argued that the bimodality could also be an artifact due to incorrect inclination-corrections applied to the $\mu_0$ values. They noted that the bimodality could also result from small number statistics. \citet{2009MNRAS.394.2022M} studied a larger number of galaxies in the Virgo cluster to overcome the problem of small number statistics and found a bimodal distribution too. However, they were hesitant to claim that bimodality is inherent to all environments because they had studied only one cluster and felt that different environments could show different behavior of $\mu_0$.\\

The goal of this paper is to address all of these issues with a study of 438 galaxies selected from the {\it Spitzer Survey of Stellar Structure in Galaxies} \citep[$S^4G$,][]{2010PASP..122.1397S}.  We study the $\mu_0$ distribution to confirm the evidence of the gap found in the Ursa Major and Virgo clusters. The confirmation of a gap can place constraints on present day galaxy distributions which are essential to the comprehension of galaxy formation and evolution. Post-Basic Calibrated Data (PBCD) used for measurements are publicly available at the Spitzer Heritage Archive website. We show that the 438 galaxies constitute a representative sample to an absolute magnitude of -15 in the 3.6 $\mu m$ band of the Spitzer Space Telescope \citep{2004ApJS..154....1W, 2004ApJS..154...10F}.  The sample extends up to 20 Mpc and includes all morphological types later than $S0^-$. The key attributes of the sample are that it is a homogeneous dataset with a large number of galaxies from the field and cluster environments imaged with excellent photometry \citep{2005PASP..117..978R}.  The mid-infrared wavelengths also offer a view of galaxies at very low extinction \citep{1984ApJ...285...89D} and with 4 minutes of integration time per pixel, the data are significantly deeper than anything that can be obtained from the ground for a large sample of galaxies.\\ 

We discuss the sample selection, $\mu_0^{[3.6]}$ measurements and corrections in the first two sections. In the next section, we discuss the bimodal distribution of $\mu_0^{[3.6]}$. In the last section, we test the likelihood of obtaining a dip at ISBs from statistical fluctuations of a flat $\mu_0$ distribution selection.  \\ 


\section{Sample Selection}

The $S^4G$ survey is a volume-, magnitude-, and diameter-limited ($d<40\;Mpc$, $|b|>30^o$, $m_{Bcorr}<15.5$ and $D_{25}>1'$) survey of over 2,350 galaxies observed with channels 1 and 2 (3.6 and 4.5 $\mu m$ respectively) of the IRAC instrument \citep{2004ApJS..154...10F} aboard the Spitzer Space Telescope \citep{2004ApJS..154....1W}. It is a very large extremely deep, representative and homogeneous sample of nearby galaxies containing all Hubble types.\\

\begin{figure}
\centering
\hspace{1em}\includegraphics[scale=0.58]{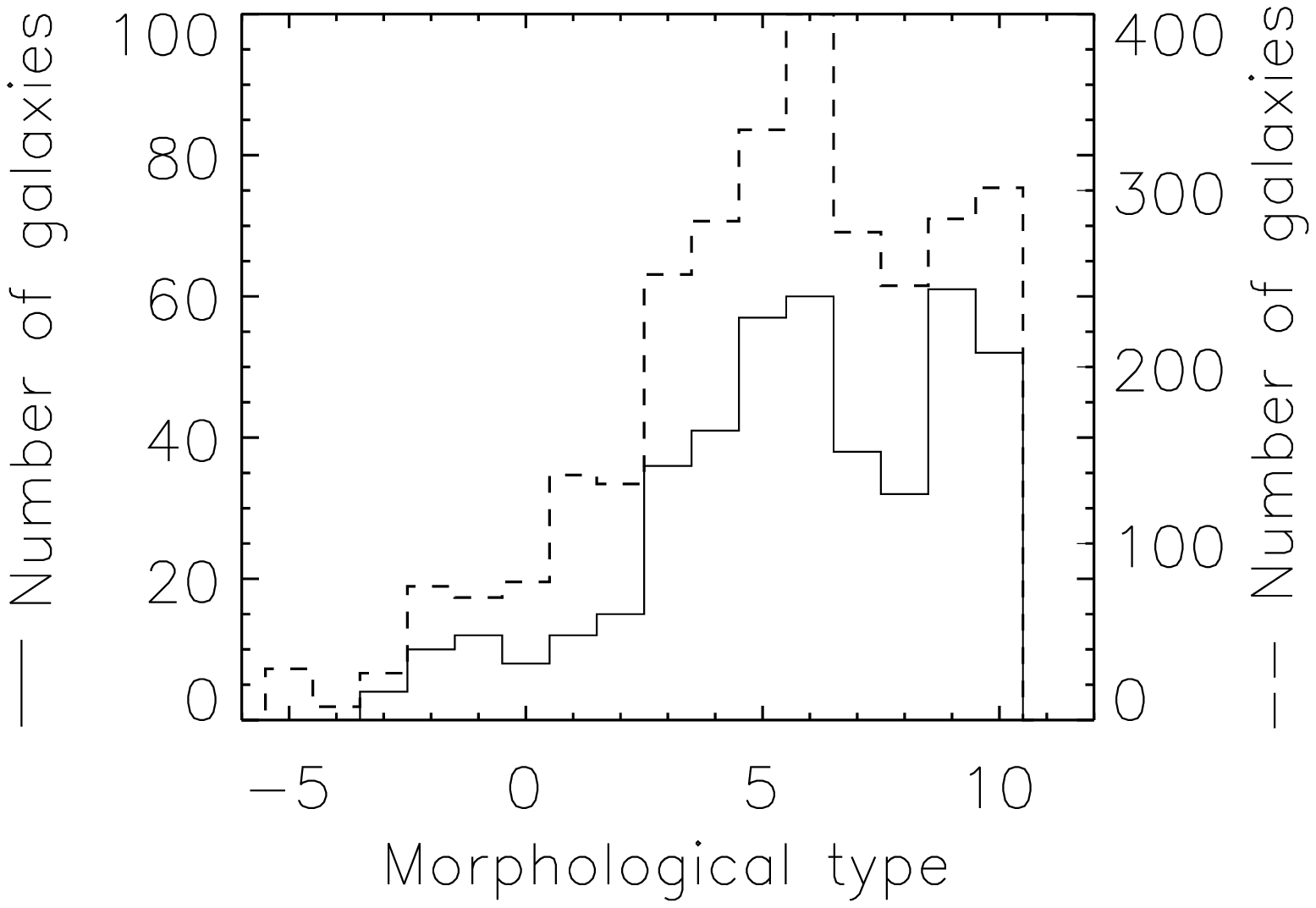}\\
\hspace{-3em}\includegraphics[scale=0.58]{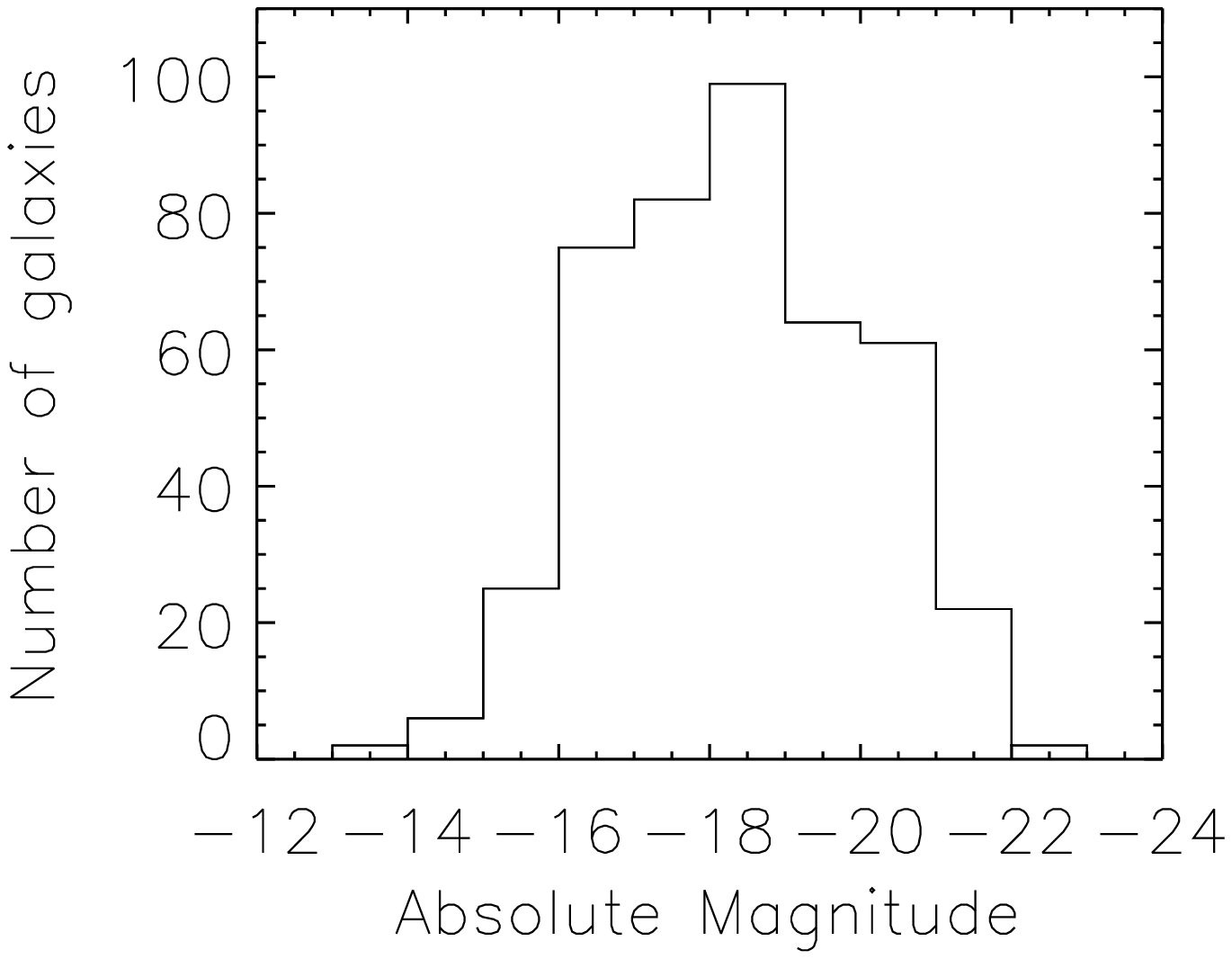}\\
\hspace{-3em}\includegraphics[scale=0.5]{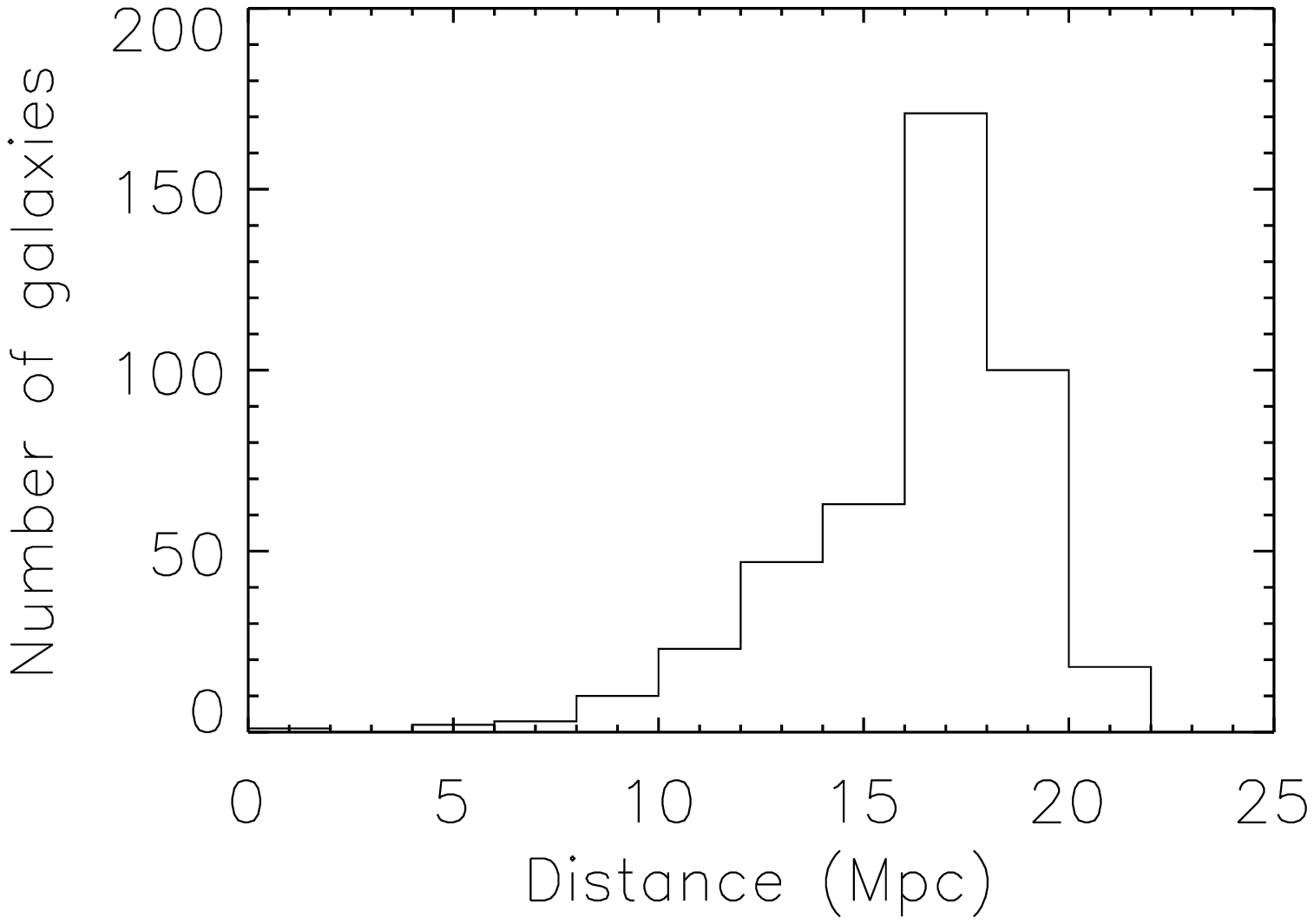}
\caption{Histograms showing the distributions of the 438 galaxies extracted from the $S^4G$ sample. Top: Morphological type from HyperLeda database (we chose all galaxies with T $>$ -3 or later than $S0^-$).  The dashed histogram is the distribution of HyperLeda types for the whole $S^4G$ sample. Middle: Absolute Magnitude (M$_{[3.6]} < -16$ in the AB system). Bottom: Distance to the galaxies from the Extragalactic Distance Database.}
\label{type-absMag-dis}
\end{figure}

\begin{figure}
\centering
\includegraphics[scale=0.55]{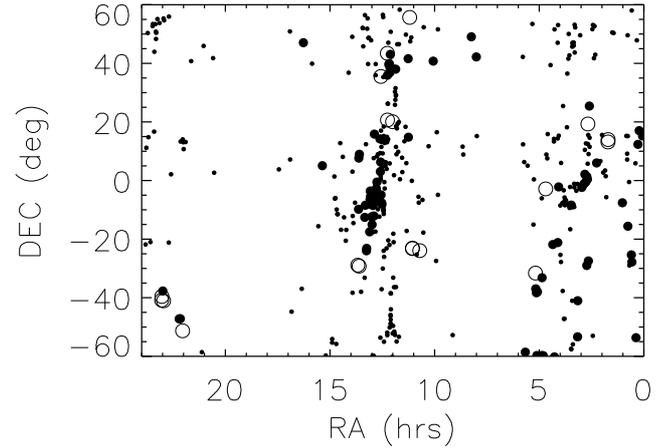}
\caption{Angular distribution on the sky of the 438 $S^4G$ galaxies. Size of the symbols shows the distance to the galaxy. The largest symbols show galaxies closer than 10 Mpc, the medium sized symbols show galaxies between 10 and 15 Mpc, and the smallest symbols show galaxies farther than 15 Mpc.}
\label{position}
\end{figure}

We use only the 3.6 $\mu m$ band data already selected before \citep[e.g.][]{2013ApJ...765...94S,2012ApJ...758L..12S} for the Cosmic Flows project \citep[e.g.][]{2012ApJ...744...43C,2012ApJ...749..174C,2012ApJ...749...78T}. The preference for the [3.6] band over the [4.5] band is based on the knowledge that 4.5 $\mu m$ fluxes have a higher contribution from hot dust than fluxes at 3.6 $\mu m$  \citep{2012AJ....144..133S,2012ApJ...744...17M}. We extracted from the $S^4G$ survey every galaxy of type later than $S0^-$ up to a distance of roughly 20 Mpc according to the Extragalactic Distance Database \footnote{http://edd.ifa.hawaii.edu} \citep[EDD,][]{2009AJ....138..323T}. This database gathers galaxy catalogs and distance measurements from several methods such as Tip of the Red Giant Branch \citep{1993ApJ...417..553L,2006AJ....132.2729M,2007ApJ...661..815R}, Cepheids' period-luminosity \citep{2001ApJ...553...47F,2011AJ....142..192F}, surface brightness fluctuation \citep{2012Ap&SS.341..179B} and Tully-Fisher \citep{1977A&A....54..661T} relations. At low redshift, surface brightnesses are independent from distances. Thus, distances used here are simple estimates derived from redshifts tethered to a Virgo infall model constrained by distance measurements. The resulting sample goes down to an absolute magnitude limit of -16 in the B band in the Vega magnitude system. This faint magnitude limit prevents the loss of low surface brightness galaxies \citep[LSB,][]{2008MNRAS.391..986Z} from the volume surveyed and guarantees the presence of galaxies of intermediate surface brightnesses. Figure \ref{type-absMag-dis} shows histograms of the sample as function of morphological type, absolute 3.6 $\mu m$ magnitude and distance. The top panel in Figure \ref{type-absMag-dis}  shows a deficit of Type-7/8 galaxies and a small excess of Type-5/6 galaxies but this is not a bias from the selection -- the distribution of the full $S^4G$ sample shown with the dashed line also displays the same behavior. Moreover, the morphological T-type assignments from HyperLeda \citep{2003A&A...412...45P} are qualitative with an uncertainty of $\delta$T=1.  There is also no a priori expectation of similar numbers of galaxies in each category in a given volume.  Figure \ref{position} shows the angular distribution across the sky of the 438 galaxies; point sizes are set according to their distances from us. \\

\section{Surface Photometry}

\subsection{Disk-only fits}

\begin{figure}
\centering
\includegraphics[scale=0.5]{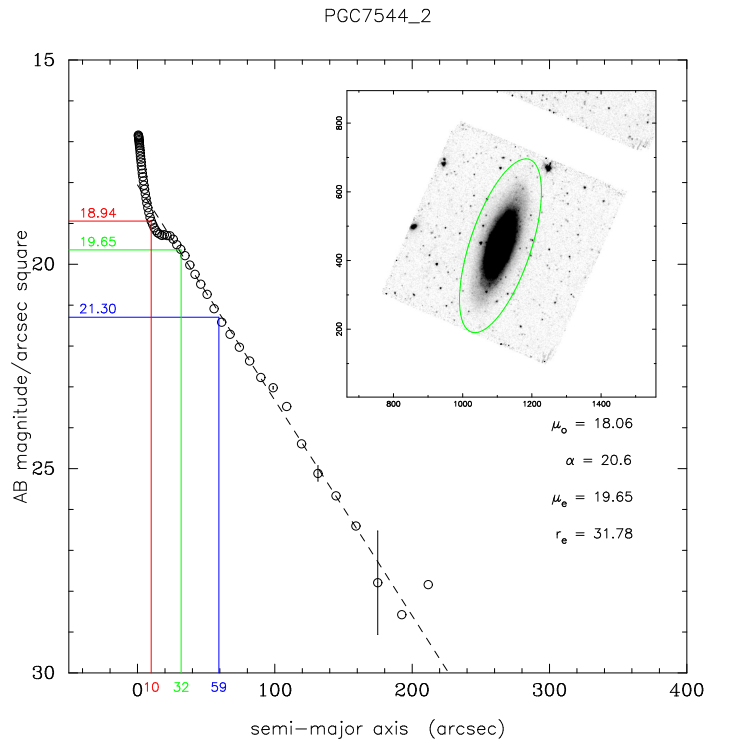}
\caption{Example of an Archangel output surface brightness profile and its disk-only fit between $a_e$ and $a_{26.5}$ at 3.6 microns. Parameters of the fit, $\mu_0^{[3.6]}$ ($\mu_0$) and the scale length ($\alpha$) are listed on the figure. Two more parameters also listed are the effective surface brightness ($\mu_e$) and radius ($r_e=a_e$). No correction has been applied yet to the data.The red, green and blue lines show the radii enclosing 20\%, 50\% and 80 \% of the total light. The green ellipse on the inset image  represents the 26.5 $mag\;arcsec^{-2}$ isophote at 3.6 microns.}
\label{sb}
\end{figure}

We perform aperture photometry for the 438 PBCD of the galaxy sample using the Archangel software developed by \citet{2007astro.ph..3646S,2012PASA...29..174S} following the procedure in \citet{2012AJ....144..133S}. Surface brightness profiles as a function of radius are fitted by simple straight lines between $a_e$ (radius of the isophote encompassing half of the total light in the [3.6] band) and $a_{26.5}$ (radius of the 26.5 $mag\;arcsec^{-2}$ isophote in the [3.6] band). In rare cases, fits are adjusted by eye, if they are clearly inappropriate between $a_e$ and $a_{26.5}$ after checking that the background brightness variation is not causing any unexpected surface brightness profile changes. Figure \ref{sb} displays an example of surface brightness fit obtained with Archangel.  $\mu_0$ is obtained by extrapolation as shown by  the following relationship: \\

\begin {equation}
\mu^{[3.6]}(r)=\mu_0^{[3.6]} + 1.0857 \frac{r}{\alpha}
\end{equation}

This corresponds to an exponential profile $L^{[3.6]}(r)=L_0^{[3.6]}e^{-r/\alpha}$. $L_0^{[3.6]}$ and $\mu_0^{[3.6]}$ are the disk central surface brightnesses in intensity and magnitude units respectively. $\alpha$ is the disk scale length.\\

\citet{1997ApJ...484..145T} and \citet{2009MNRAS.393..628M}  were concerned about the disk-only fitting technique for bulge galaxies. However, they respectively showed that neither dropping bulge galaxies nor making a bulge-disk decomposition removed the bimodality. They also asserted that their results were not significantly affected by alternative decompositions.  Fitting galaxy profiles according to their structures (bulges, bars) allows for an interplay between components that can conceal the global properties that concern this study.  We use the consistent disk-only fitting technique without exception across the entire sample. \citet{1996A&AS..118..557D} showed that disk-only fits gave unbiased disk parameters relative to a 2D fit decomposition parameters within 0.5 $mag\, arcsec^{-2}$. Since the background (distant galaxies and zodiacal light) uncertainties lead to a small magnitude uncertainty even for IRAC ch1 \citep{2012AJ....144..133S}, we retain an  uncertainty budget of 0.5 $mag\,arcsec^{-2}$ for $\mu_0^{[3.6]}$ measurements. This leads then to a choice of bin sizes of 0.5 $mag\,arcsec^{-2}$ for histograms of $\mu_0^{[3.6]}$  distributions.

\subsection{Corrections}

\subsubsection{Aperture Correction}

The photometry for the IRAC instrument is normalized to a finite aperture (12") which requires aperture corrections  (\citet{2005PASP..117..978R} and IRAC Instrument Handbook) for large extended sources such as galaxies. The following correction\footnote{http://irsa.ipac.caltech.edu/data/SPITZER/docs/irac/iracinstrumenthandbook/}  recommended for fluxes is applied to measurements: 

\begin{equation}
F_{corr}=F_{mes}\times (A \times exp (-r^B) + C)
\label{apercorr}
\end{equation}
where $F_{corr}$ is the flux corrected for aperture, $F_{mes}$ is the measured flux, r is the aperture radius in arcseconds. The correction is very small for IRAC ch.1 where A=0.82, B=0.37 and C=0.91. For example, the disk central surface brightness of PGC7544 shown in Figure \ref{sb} becomes 18.10 after correction against 18.06 before. This is within the retained uncertainty on disk central surface brightness measurements. Hereafter, $\mu_0^{[3.6],a}$ designates the disk central surface brightness corrected for aperture. It results from the fit of the corrected surface brightness profile.

\subsubsection{Inclination Correction}

Inclination effects are quite confusing at optical band because the path length of observed surface brightnesses varying with the line-of-sight, and extinction work in opposite way on measured $\mu_0$ values.  At 3.6 microns we can assume the obscuration to be negligible and therefore only the geometric effect of the inclination on surface brightnesses needs to be taken into account.  We measure the face-on $\mu_0^{[3.6],a}$ as follows:

\begin{equation}
\mu_0^{[3.6],a,i} = \mu_0^{[3.6],a}-2.5C^{[3.6]} log (\frac{b}{a})
\label{inccorr}
\end{equation}
with $C^{[3.6]}$=1 assumed for a transparent system. $\mu_0^{[3.6],a,i}$ is the disk central surface brightness corrected for both aperture and inclination, and $\mu_0^{[3.6],a}$ is the disk central surface brightness corrected for aperture corrections above (Eqn. \ref{apercorr}).\\

\subsection{The Ursa Major Cluster}
\label{UMasection}

\begin{figure}
\centering
\includegraphics[scale=0.5]{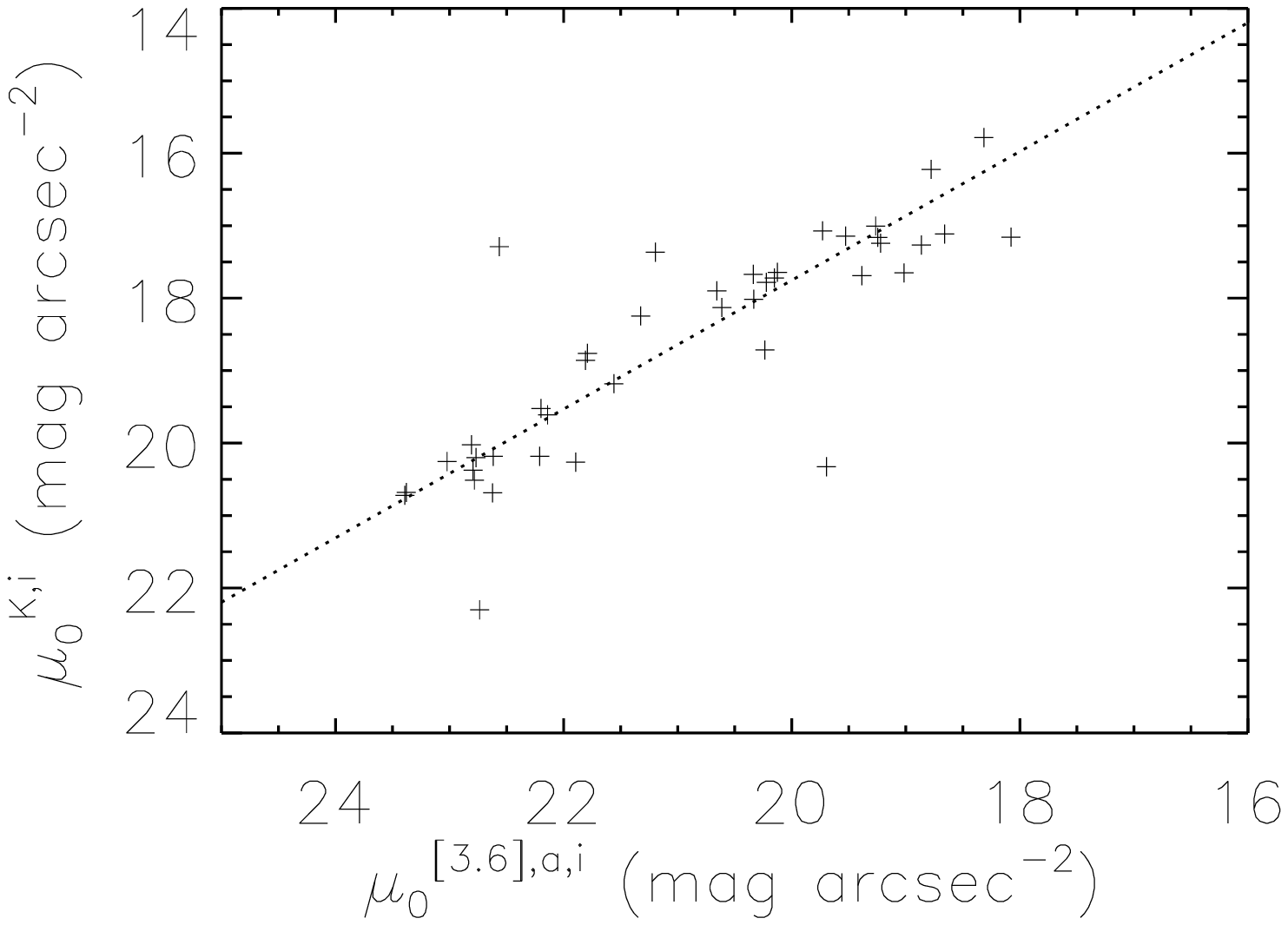}\\
\vspace{0.3cm}
\includegraphics[scale=0.5]{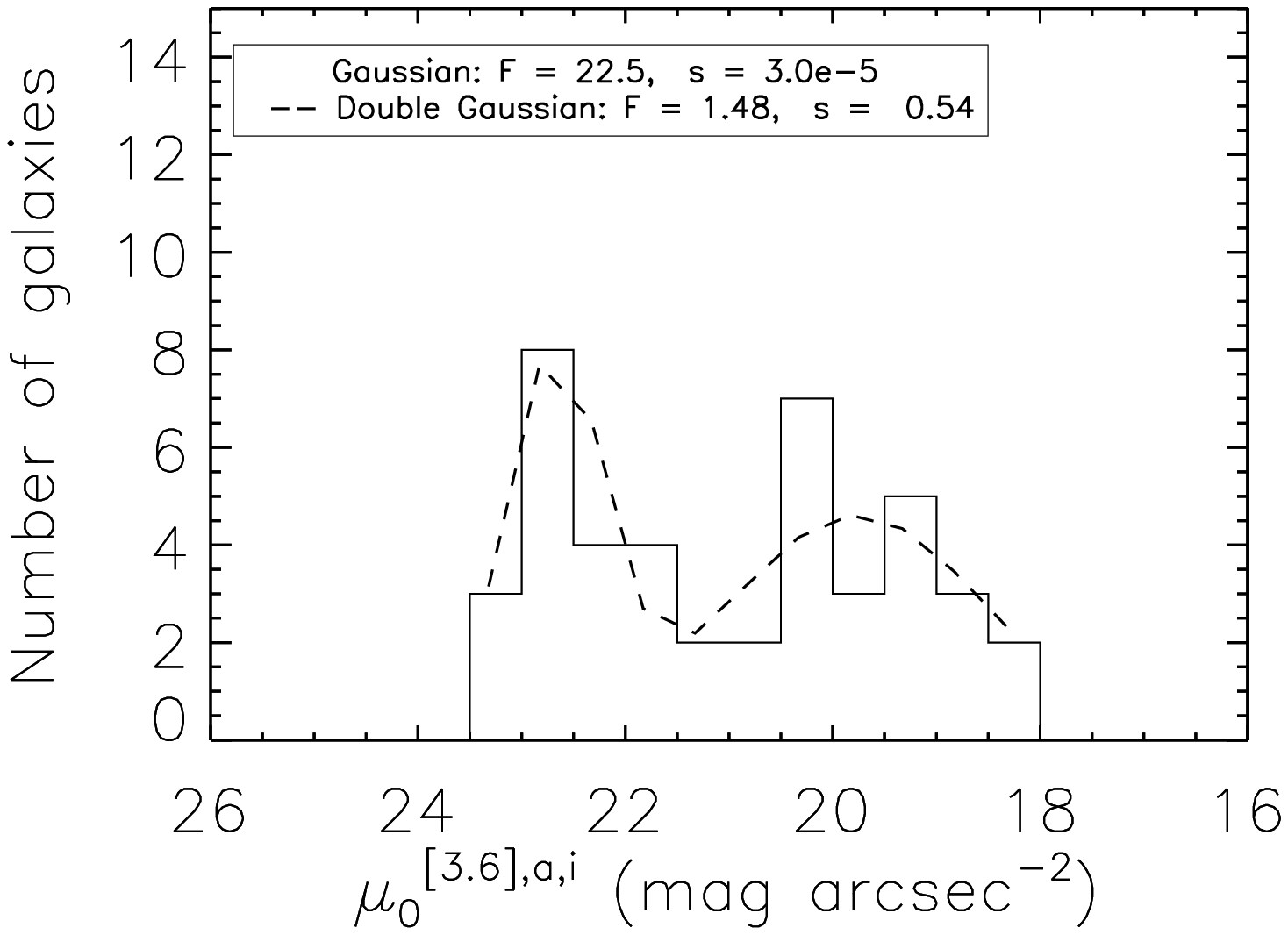}\\
\vspace{0.3cm}
\includegraphics[scale=0.5]{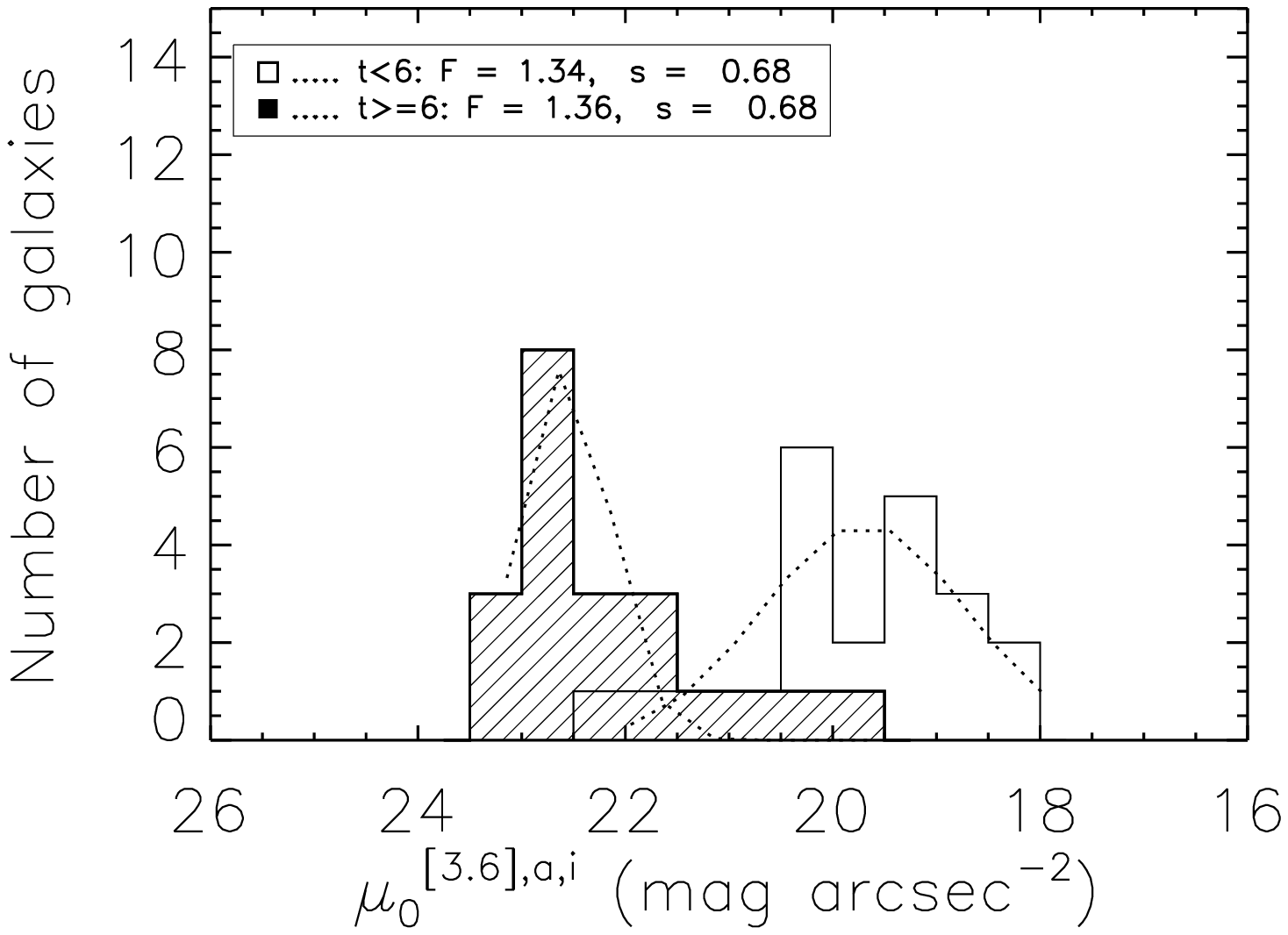}
\caption{Top: Comparison between disk central surface brightnesses from Tully \& Verheijen 1997 in the K'-band, Vega system and this paper, AB system. The dotted line is the 1:1 relation shifted by the average K'(Vega)-[3.6](AB) value of the sample. Middle: Histogram of the disk central surface brightnesses corrected for inclination and aperture of 43 Ursa Major galaxies. The dashed line represents the sum of two gaussian fits from LSB and HSB subsample modelings. While the significance level (s) of the F-test (F) for the double gaussian is high, the single gaussian modeling can be rejected at more than the 99\% confidence level. Bottom: Distributions of the type greater than 6 galaxy disk central surface brightnesses (dashed histogram) and type smaller than 6 galaxy disk central surface brightnesses (plain histogram) for the available galaxies of the Ursa Major cluster.}
\label{UMa}
\end{figure}

To test the measurements of $\mu_0$ at 3.6 $\mu m$ to that obtained in previous papers, we retrieve the available 43 of 78 previously used Ursa Major galaxies \citep{1997ApJ...484..145T} from the Spitzer archive.  16 of these galaxies are from the Carnegie Hubble Program \citep{2011AJ....142..192F} and 27 are from the $S^4G$ survey. $\mu_0$ values obtained with our procedure at 3.6 microns in the AB system are compared to values obtained in the K' band in the Vega system from the 1997 paper in the top panel of Figure \ref{UMa}. Figure \ref{UMa} top shows a good agreement between K'band values and [3.6] band values.\\

The $\mu_0$ distributions are shown in the middle and bottom panels of Figure \ref{UMa} for the 43 galaxies in common with the previous studies. The bottom panel also shades the galaxies according to their type - one immediately sees that the early type galaxies have a higher $\mu_0$ (ie, are usually HSBs) compared to the usually lower surface brightness late type galaxies demonstrating the known strong correlation between morphological type and the disk central surface brightness. The previous studies found values for LSB and HSB peaks in the K'-band at about 19.7--20.0 for LSBs and 17.3--17.5 for HSBs ($mag\;arcsec^{-2}$ in the Vega system). We find values of 22.5 and 19.5 for LSBs and HSBs at 3.6 $\mu$m but in the AB system as shown in Figure \ref{UMa}. A conversion between the Vega and AB systems re-establishes the proper relative positions between peaks in the K' and [3.6] bands. Namely, at 3.6 microns, peaks are located at smaller values in the Vega system ($[3.6](Vega) = [3.6](AB) - 2.785$) \citep{2012AJ....144..133S} than in the K' band in the same system. A single gaussian model of the $\mu_0$ distribution can be rejected at more than the 99\% confidence level while the significance of a double gaussian modeling is quite high (54\%). However these 43 galaxies do not constitute a complete sample and cannot lead to a universal conclusion about the bimodality we seek to test. Still the comparisons show that our procedure gives results similar to the previous studies and does not introduce any particular bias.\\

\section{Results}

\subsection{A Lack of Intermediate Surface Brightness galaxies}

\begin{figure}
\centering
\includegraphics[scale=0.5]{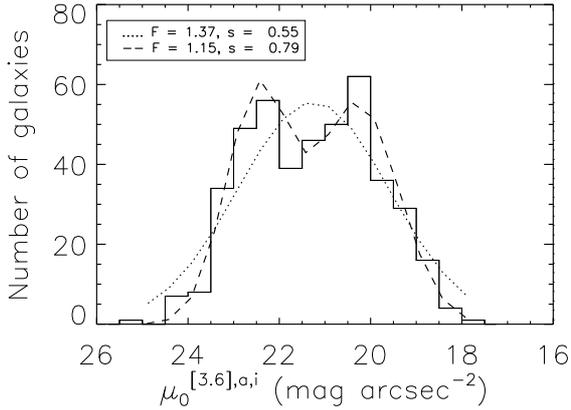}
\caption{Histogram of the aperture-inclination corrected disk central surface brightnesses. A gaussian fit (dotted line) to the distribution can be rejected at the 55 \% confidence level (F=F-Test, s=significance level) while a sum of two gaussians respectively from the modelings of LSB and HSB disk central surface brightness subsample distributions can be rejected only at the 21\% confidence level (dashed line) supporting the bimodality seen previously.}
\label{mui}
\end{figure}

\begin{figure}
\centering
\includegraphics[scale=0.55]{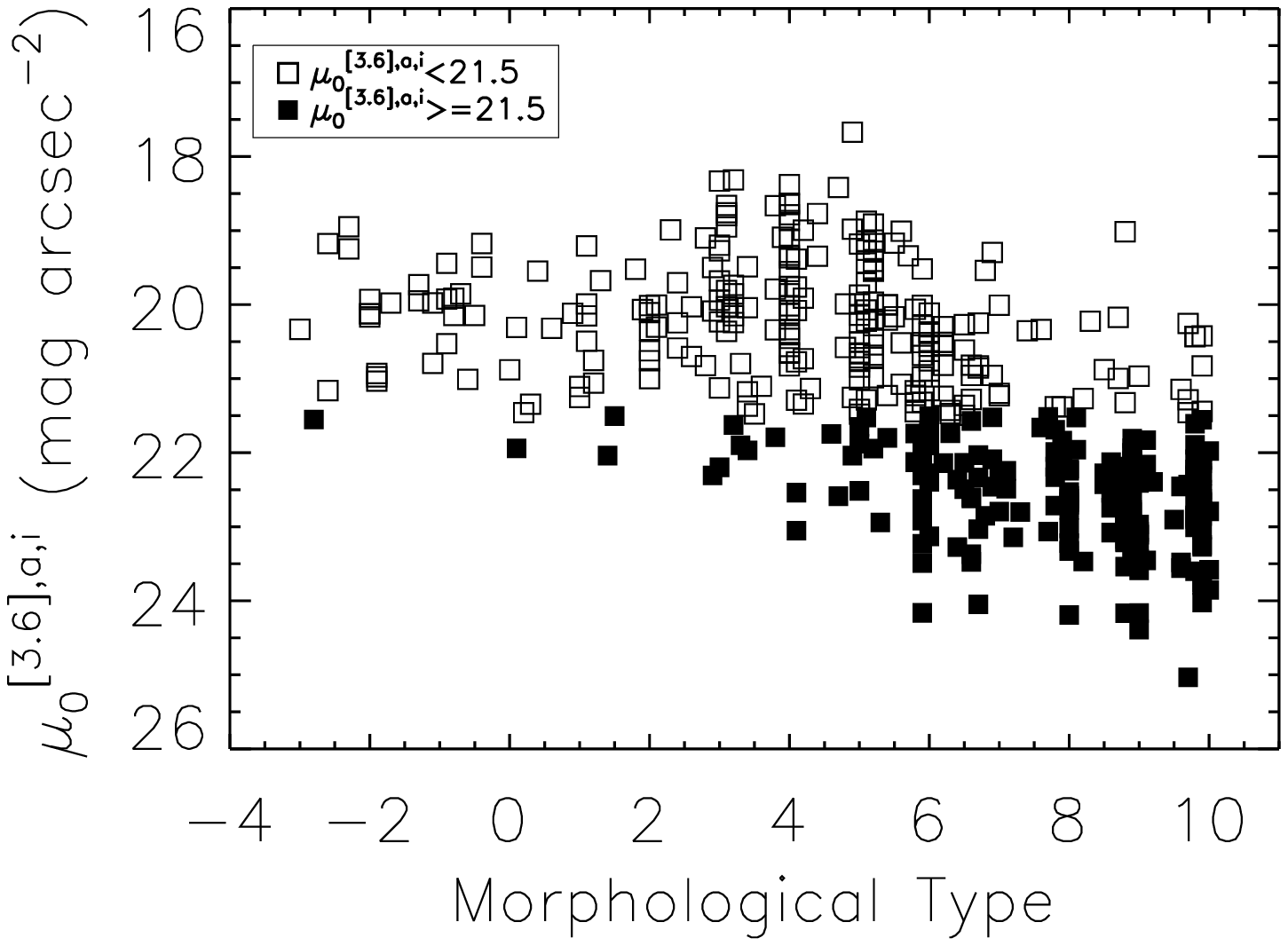}
\hspace{-2.5em}\includegraphics[scale=0.55]{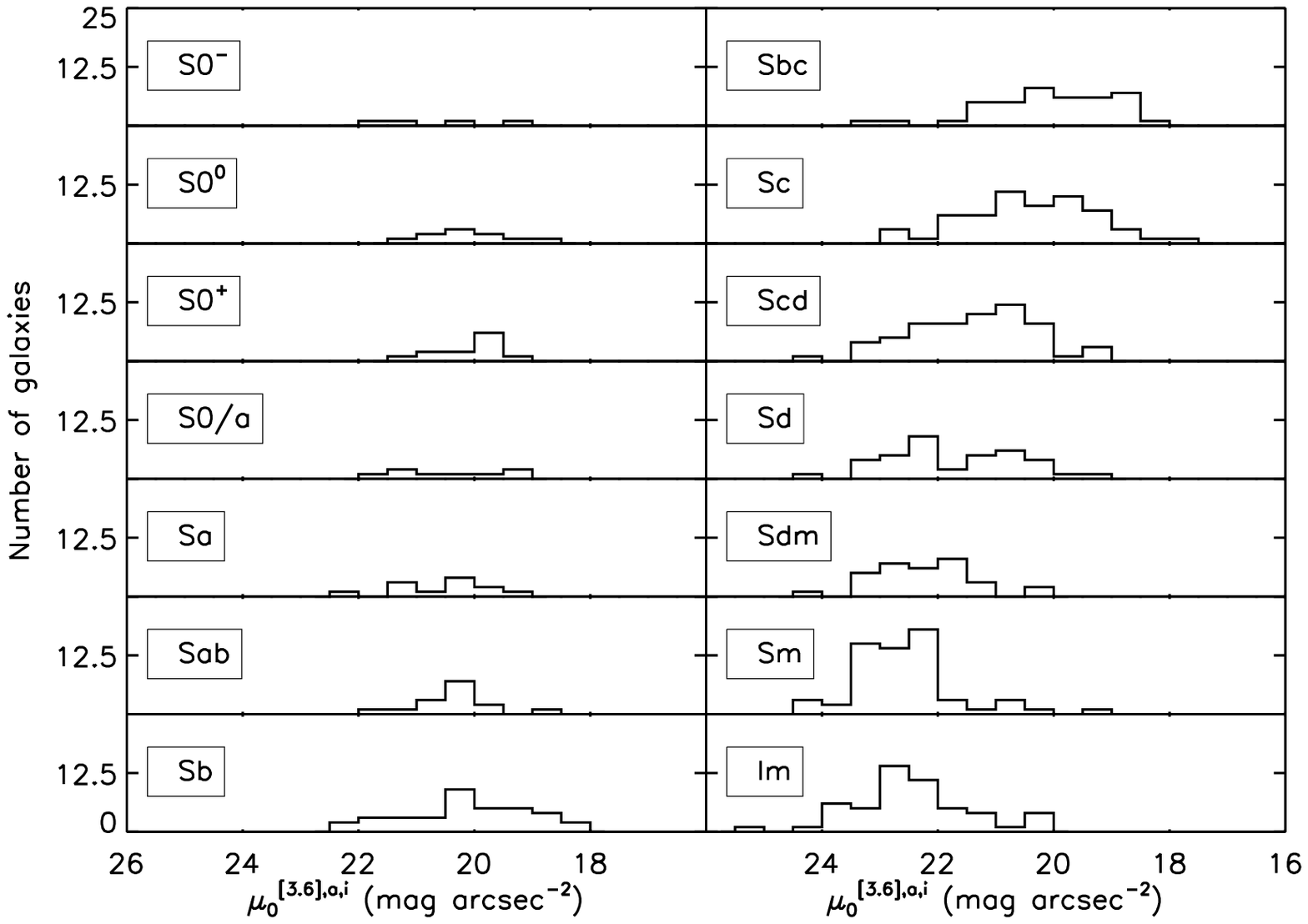}
\caption{Top: disk central surface brightness versus morphological type. An ad-hoc limit set at $\mu_0^{[3.6],a,i}=21.5\, mag\,arcsec^{-2}$ shows that galaxies with $\mu_0^{[3.6],a,i}>21.5\, mag\,arcsec^{-2}$ appear for types greater than $\sim$ 6 (later than Scd). Bottom: Peaks of the $\mu_0^{[3.6],a,i}$ distributions shift to the left with increasing morphological type.}
\label{typemui}
\end{figure}

 The entire $\mu_0^{[3.6],a,i}$ distribution for the 438 galaxy sample from S$^4$G is shown in Figure \ref{mui}.  It reveals a hint of bimodality  -- There is a lack of galaxies between $21\;mag\;arcsec^{-2}$ and $22\;mag\;arcsec^{-2}$ (less than 45 galaxies versus more than 60 galaxies at the peaks).  We fit this distribution with a double gaussian assuming a population with $\mu_0$ greater than 21.5 $mag\;arcsec^{-2}$ and a population with $\mu_0$  less than 21.5 $mag\;arcsec^{-2}$ - roughly reflecting a LSB and HSB population respectively (see top panel in Figure \ref{typemui}) .  This model has a much higher significance (can only be rejected at a 21\% confidence level) compared to a simple gaussian model which can be rejected at a 55\% confidence level. \\

As shown in the top panel of Figure \ref{typemui}, the ad-hoc limit of $21.5\;mag\;arcsec^{-2}$ matches well with the well known fact that LSB galaxies are in general of late Hubble type while HSB galaxies are in general of early Hubble type \citep[e.g.][]{1996A&A...313...45D}.  In this figure, the LSB galaxies appear only for types later than the Scd type.  The bottom panel of Figure \ref{typemui} shows the $\mu_0^{[3.6],a,i}$ distribution for each morphological type in the 438 galaxy sample.  The histogram peak shifts to the left (higher $\mu_0^{[3.6],a,i}$) with increasing morphological types, supporting the correlation between types and the disk central surface brightnesses.

\subsection{Selection criteria}

Since the double gaussian modeling can be rejected as a model for the total sample at the 21\% (against 55\% for the single gaussian modeling) confidence level, we next define selection criteria to understand better which galaxies are in the gap and what might be causing a bimodality. 

\subsubsection{Close Neighbor}

Galaxies with a close neighbor ($d<80\;kpc$) can undergo interactions which can potentially modify their $\mu_0$ values. If there are only two parent populations (HSB and LSB types), then mergers / interactions could move galaxies into the gap separating the HSB and LSB galaxies \citep{1997ApJ...484..145T}. We remove galaxies with close neighbors to test the postulate that this will increase the bimodal nature of the $\mu_0$ distribution.\\

\begin{figure}
\centering
\hspace{0.1em}\includegraphics[scale=0.5]{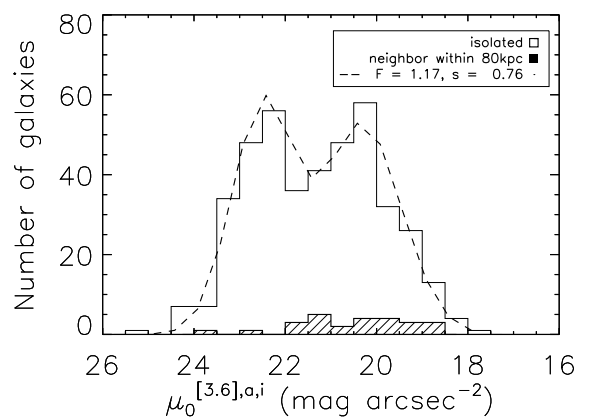}\\
\caption{Separation between galaxies with at least one neighbor closer than 80 kpc (dashed histogram) and without a neighbor within 80 kpc (plain histogram). The velocity of the neighbor has to be within $\pm$ 200 $km\;s^{-1}$ of the galaxy velocity.
The dashed line is the sum of two gaussians. It cannot be rejected at more than the 24 \% confidence level (1-significance level (s)) with a F-test (F).}
\label{multiple}
\end{figure}

We assume that a galaxy with a velocity, v, has a "close neighbor" whenever there is another galaxy within 80 kpc with a velocity equal to v$\pm$ 200 $km\;s^{-1}$.  We find only a few galaxies with close neighbors as shown on Figure \ref{multiple}.  The remaining "isolated" sample consists of 411 galaxies. \\

The distribution of the 27 galaxies with close neighbors looks quite flat - although the numbers are too small to investigate the nature of the distribution. Quite a few of the 27 galaxies are in the gap. This agrees with the hypothetical scenario proposed by \citet{1997ApJ...484..145T} in which LSB galaxies tend to turn into HSB galaxies progressively going through a stage as intermediate surface brightness galaxies due to an interaction with a neighbor.

\subsubsection{Inclination}

Next we explore the effect of inclination corrections as noted by \citet{2000MNRAS.311..668B}. According to them, the effects of dust and projection geometry may not be negligible, even in the mid-infrared: 1) averaging ellipse surfaces at high inclinations to obtain $\mu_0$ may result in a systematically smaller value \citep{1994essg.book.....H}, 2) assuming a thin, uniform, slab disk at high inclinations may lead to incorrect conclusions because three dimensionality of stellar structures affects the inclination correction in non-trivial ways.  It is difficult to characterize the surface brightnesses of edge-on galaxies  because integrating along the line of sight may hide the effects of sub-structures like bars and spiral arms \citep[e.g.][]{2010MNRAS.401..559M}.  No precise method exists for correcting such effects and the coefficient $C^{[3.6]}$ in equation \ref{inccorr} itself may vary with galactic radius. In absence of a correct (or even better) method to correct for inclination, we decided to remove every galaxy with an inclination greater than $\sim73^o$ ($\frac{b}{a}$=0.35)\footnote{0.4 instead of 0.35 was the choice of \citet{2000MNRAS.311..668B}. Because of the uncertainties on inclinations ($\sim$ $4-5^0$), choosing 0.35 ($73^0$) over 0.4 ($69^0$) does not change the conclusions.}  leaving us with 292 galaxies. \\

\begin{figure}
\centering
\includegraphics[scale=0.5]{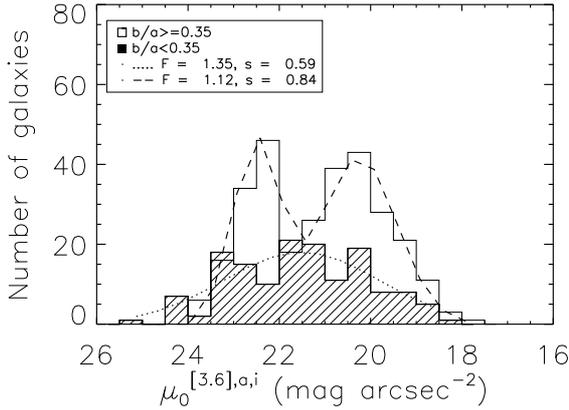}\\
\caption{Distribution of disk central surface brightnesses of highly inclined galaxies ($i>73^o$, dashed histogram) against the others ($i<73^o$, plain histogram). The $\mu_0^{[3.6],a,i}$ distribution of highly inclined galaxies is fitted by a gaussian (dotted line) whereas the less inclined galaxy $\mu_0^{[3.6],a,i}$  distribution is fitted by a double gaussian (dashed line) (F=F-Test, s=significance level).}
\label{inc}
\end{figure}

Figure \ref{inc} displays the bimodal distribution of this refined sample along with the distribution for the "edge-on" galaxies.  The inclined galaxies are well-described by a single gaussian that peaks in the previously observed gap between the HSB and LSB galaxies.   The remainder of the sample is well-fit with a double gaussian model which cannot be rejected at more than the 16 \% confidence level with the F-Test.

\subsubsection{Axial Ratio}

\begin{figure}
\centering
\hspace{0.1em}\includegraphics[scale=0.5]{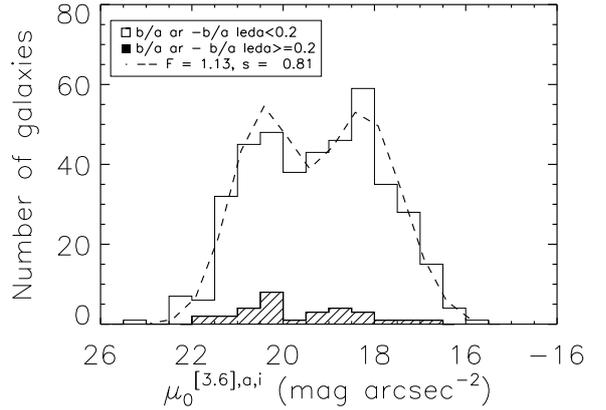}\\
\caption{Distribution of $\mu_0^{[3.6],a,i}$ separating galaxies which ''Archangel axial ratios'' differ by more than 0.2 from ''HyperLeda axial ratios'' (dashed histogram) from which that do not (plain histogram). The total sample is modeled by a sum of two gaussians (dashed line) (F=F-Test, s=significance level).}
\label{diff}
\end{figure}

An error in the axial ratio is another source of error in the inclination correction we apply to get the face-on $\mu_0^{[3.6],a}$ value. An error of 0.2 in the axial ratio leads to an error of about 0.5 $mag\;arcsec^{-2}$.  Archangel axial ratios are the means of the axial ratios of the fitted ellipses to galaxies between 50\% and 80\% of their total luminosities.  We remove every galaxy whose Archangel derived ratios differ from that found in HyperLeda database \citep{2003A&A...412...45P} by more than 0.2.  The resulting distribution of $\mu_0^{[3.6],a,i}$ is shown in Figure \ref{diff}.   Where the axial ratio between the HyperLeda values and that measured by Archangel in the S$^4$G data are similar  (408 galaxies), the bimodal distribution is still visible. There are fewer LSB galaxies in this plot compared to the original because LSB galaxies are the ones most affected by a change in observations (optical versus mid-infrared). The double gaussian modeling cannot be rejected at more than the 19\% confidence level.

\subsubsection{Combination of selection criteria}

\begin{figure}
\centering
\hspace{0.05em}\includegraphics[scale=0.5]{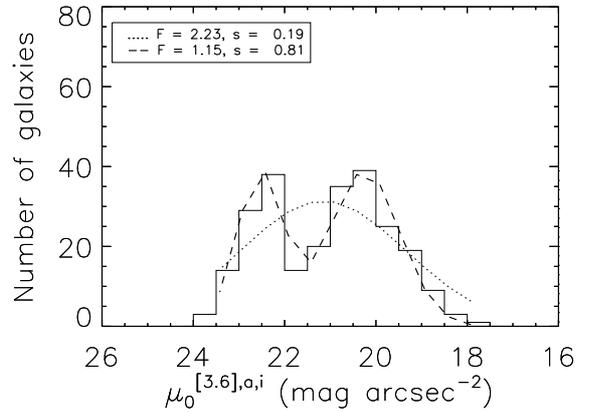}\\
\caption{$\mu_0^{[3.6],a,i}$ histogram after dropping galaxies inclined by more than $73^o$, with a neighbor closer than 80 kpc and those with an axial ratio different by more than 0.2 between HyperLeda and S$^4$G data. A double gaussian (dashed line) clearly fits the distribution far better than a single gaussian (dotted line) (F=F-Test, s=significance level).}
\label{all}
\end{figure}

If we now combine the three selection criteria (isolated, non-inclined galaxies, with similar axial ratios in the optical and mid-infrared)  we get a sample of 249 galaxies.  Figure \ref{all} shows the $\mu_0^{[3.6],a,i}$ distribution for this highly refined sample and clearly shows the bimodality. The double gaussian modeling now cannot be rejected at more than the 19\% confidence level according to the F-Test. On the contrary, the single gaussian modeling can be rejected at a 81\% confidence level with the same test. Thus we conclude that the $\mu_0^{[3.6],a,i}$ distribution of the sample is bimodal and that there is a lack of intermediate surface brightness galaxies.\\

\section{Discussion}

\subsection{Likelihood of getting a dip at Intermediate Surface Brightness}

The different results of previous sections demonstrate strongly that the $\mu_0^{[3.6],a,i}$ distribution is bimodal. Is it possible that the bimodality is from statistical fluctuations? To obtain the likelihood of getting a dip at intermediate surface brightness from a statistical fluctuation, we simulated a flat $\mu_0^{[3.6],a,i}$ distribution of 700 galaxies with $\mu_0$ between 19 and 24 $mag\;arcsec^{-2}$. We chose a flat distribution in agreement with the current description of disk central surface brightness distribution \citep{1995AJ....110..573M,1996MNRAS.280..337M}. This flat distribution has an upper brightness limit to disk central surface brightnesses that must have a physical origin \citep{1970ApJ...160..811F} and a lower brightness limit due to observational limitations in the simulations. Each 1 $mag\;arcsec^{-2}$ size bin contained the same number of galaxies. We randomly selected 249 galaxies from that distribution and looked at the likelihood of obtaining a gap between two peaks. We repeated the selection 10,000 times and retained only simulations with a gap between two peaks and a number of galaxies in the gap no greater than 50 \% the number of galaxies in the peaks. This is approximately what we observe in Figure \ref{all}. Table \ref{table1} lists the histogram parameters of the 9 simulations out of the 10,000 simulations which showed a distribution similar to the observed bimodality we observed in the 249 galaxy sample. The last line corresponds to the mean values for the 9 simulations. Thus the likelihood of randomly obtaining the observed bimodality is only $\sim$ 0.1 \%.

\begin{table*}
\begin{center}
\begin{tabular}{|c|c|c|c|c|c|c|c|c|}
Simulations & peak1 & position peak1 & peak2 & position peak2  & gap & position gap & $\%_1$ & $\%_2$ \\
\hline
   431 & 71 &  24.0 & 62 &  22.0 & 31 &  23.0 &  44 &  50\\
   870 & 63 &  23.0 & 61 &  21.0 & 27 &  22.0 &  43 &  44\\
  2603 & 64 &  24.0 & 63 &  21.0 & 30 &  22.0 &  47 &  48\\
  3923 & 65 &  20.0 & 59 &  23.0 & 28 &  22.0 &  43 &  48\\
  4852 & 66 &  24.0 & 61 &  22.0 & 23 &  23.0 &  35 &  38\\
  4887 & 64 &  23.0 & 63 &  20.0 & 28 &  21.0 &  44 &  44\\
  8330 & 66 &  22.0 & 62 &  24.0 & 30 &  23.0 &  46 &  48\\
  8421 & 67 &  23.0 & 60 &  20.0 & 30 &  22.0 &  45 &  50\\
  9233 & 63 &  23.0 & 62 &  21.0 & 30 &  22.0 &  48 &  48\\
  \hline
Mean of the 9 kept simulations & 65 & 23 &  61 &  21 &  28 & 22 & 44 & 46\\
\hline
\end{tabular}
\end{center}
\caption{Kept simulations: (1) Number characterizing the simulation out of 10,000. Only simulations with two peaks and a gap in between are selected. The number of galaxies in the gap has to be at most 50 \% the galaxy number in peaks. Only 9 out of 10,000 simulations are similar to the observed bimodality.  In other words, there is only a $\sim$ 0.09 \% probability that the bimodality is due to a statistical fluctuation.  (1) - (2) Value and Position of the first peak, (3) - (4) Value and Position of the second peak, (5) - (6) Value and Position of the gap, (7) - (8) Percentage of galaxies in the gap with respect to the number of galaxies in the first and second peaks respectively.}
\label{table1}
\end{table*}

\subsection{Comparisons with the literature}

In the Vega system, peaks have been found at 17.5-18 and 20 $mag\;arcsec^{2}$ for both Virgo and Ursa Major Clusters in the K' band and in the H band respectively \citep{1997ApJ...484..145T,2009MNRAS.393..628M,2009MNRAS.394.2022M}. The gap found in between has a width of one unit in magnitude per arcsecond square and is located at $\sim$19 $mag\;arcsec^{-2}$. In this study, at 3.6 $\mu m$ in the AB system, the observed bimodal distribution shows two peaks at 20.5 and 22.5 $mag\;arcsec^{-2}$ and a gap of  width 1 $mag\;arcsec^{-2}$ at 21.5 $mag\;arcsec^{-2}$ in between. This excellent agreement between the studies, including the shift to smaller values when moving towards longer wavelengths (visible after applying the AB-Vega system conversion) already shown in Figure 4 of the 1997 paper, is a strong evidence for an inherent bimodality in the local galaxy population.

\subsection{Two stable modes?}

The results reveal a clear separation between $\mu_0^{[3.6]}$ of HSB and LSB galaxies. The former are  probably dominated by baryonic matter at their centers whereas the latter are likely dark matter dominated at all radii \citep{1997ApJ...484..145T}.  Along with a lack of intermediate surface brightness galaxies, the data suggest that the two (L/HSB) peaks signify two stable configurations of galaxy formation  \citep{1963MNRAS.126..553M}.  Systems that retain large angular momentum from their formation may prevent the baryonic matter from collapsing to form a stellar disk that could dominate the dark halo at the galaxy center. These systems may reach a rotational equilibrium at densities where the dark matter halo remains dominant and the galaxies appear as LSB galaxies. On the other hand, galaxies with low angular momentum, either because of their formation or because they transferred their angular momentum away from much of their gas, allow baryonic matter to collapse and form disks that can dominate the dark matter halo at the center. This hypothesis is supported by the differences between  typical LSB and HSB rotation curves. LSB galaxies reach flat rotation at very large radii from their centers \citep{2010ApJ...718..380S} whereas HSB galaxies reach their maximal rotation speed at, or within, r=2.15$\alpha$ where $\alpha$ is the disk scale length \citep{1997AJ....114.2402C}.\\ 

The gap between the two peaks suggests that galaxies tend to avoid being in a situation where the dark matter and the baryonic matter have equal weight in the center. The few galaxies present in this gap may be transitioning from LSB to the HSB galaxies as suggested by the experiment with the close neighbor pairs.  Eventually all galaxies that undergo interactions may end up as HSB galaxies so that the peak of HSB systems should be higher than the LSB peak in environments where interactions are common. \\


\section{Conclusion}

We selected 438 galaxies from the "Spitzer Survey of Stellar Structure in Galaxies" ($S^4G$)  to measure the  disk central surface brightnesses.  The data were processed and extracted with the Archangel software and the measurements were corrected for aperture and inclination effects. The resulting $\mu_0^{[3.6],a,i}$ distribution is bimodal at the 79 \% significance level.\\

We examined the effect of three criteria (inclination, close neighbors and axial ratio measurements) on the $\mu_0^{[3.6],a,i}$ distribution. The bimodalilty did not change significantly when removing close neighbors or galaxies with significantly different axial ratios between our measurements and those in HyperLeda. However, the bimodality was most clearly evident when removing highly inclined galaxies, refuting the hypothesis that inclination-correction could be a cause of the bimodality. After removing galaxies prone to any of these three caveats, the refined sample of 249 galaxies (still 1.5--4 $\times$ larger than previous studies) showed a clear bimodality that cannot be rejected at more than a 19 \% confidence level.  The gap between the two populations is also shown not to be a statistical fluctuation at a better than 0.1\% probability level.  In particular, the excellent agreement in the position of the gap between this study and previous studies reinforces the finding of the bimodality in the $\mu_0$ distribution. The one magnitude gap between HSB and LSB populations is statistically improbable and it occurs where anticipated by prior studies with other samples.\\
  
  Our 438 galaxy sample is representative of all galaxies later than $S0^-$ in the half of the sky at the Galactic poles within the volume extending to 20 Mpc and brighter than $M_B=-16$. The galaxies lie in clusters, groups, and the field. The bimodality in disk central surface brightnesses first found in moderate and high density regions is found to be pervasive. Galaxies can be HSB or LSB but they avoid an intermediate state. This phenomenon must have a physical explanation, one that probably will give an important clue regarding the process of galaxy formation.

\section*{Acknowledgements}

We are grateful to the dedicated staff at the Spitzer Science Center for their support and help with the planning and execution of this legacy exploration program S$^4$G and the exploratory science program Cosmic Flows. This work is based on observations and archival data obtained with the Spitzer Space Telescope, which is operated by the Jet Propulsion Laboratory, California Institute of Technology under a contract with NASA. Support for this work was provided by NASA. We gratefully acknowledge support provided by NASA through two awards issue by JPL/Caltech. RBT acknowledges support from the US National Science Foundation award AST09-08846. HC and JS acknowledge support from the Lyon Institute of Origins under grant ANR-10-LABX-66. This work made use of the HyperLeda database (http://leda.univ-lyon1.fr) and of the NASA/IPAC Extragalactic Database (NED) which is operated by the
Jet Propulsion Laboratory, California Institute of Technology, under contract with the National Aeronautics and Space Administration. We thank Tom Jarrett for useful discussions regarding Spitzer data aperture correction and the anonymous referee whose comments have contributed to improve this paper.

\clearpage

\bibliography{bibliDeJong}

\begin{thebibliography}{48}
\expandafter\ifx\csname natexlab\endcsname\relax\def\natexlab#1{#1}\fi

\bibitem[{{Bailin} \& {Harris}(2008)}]{2008MNRAS.385.1835B}
{Bailin} J., {Harris} W.~E., 2008, \mnras, 385, 1835

\bibitem[{{Baldry} {et~al}\mbox{.}(2004){Baldry}, {Glazebrook}, {Brinkmann},
  {Ivezi{\'c}}, {Lupton}, {Nichol}, \& {Szalay}}]{2004ApJ...600..681B}
{Baldry} I.~K., {Glazebrook} K., {Brinkmann} J., {Ivezi{\'c}} {\v Z}., {Lupton}
  R.~H., {Nichol} R.~C., {Szalay} A.~S., 2004, \apj, 600, 681

\bibitem[{{Bell} \& {de Blok}(2000)}]{2000MNRAS.311..668B}
{Bell} E.~F., {de Blok} W.~J.~G., 2000, \mnras, 311, 668

\bibitem[{{Blakeslee}(2012)}]{2012Ap&SS.341..179B}
{Blakeslee} J.~P., 2012, \apss, 341, 179

\bibitem[{{Brammer} {et~al}\mbox{.}(2009){Brammer}, {Whitaker}, {van Dokkum},
  {Marchesini}, {Labb{\'e}}, {Franx}, {Kriek}, {Quadri}, {Illingworth}, {Lee},
  {Muzzin}, \& {Rudnick}}]{2009ApJ...706L.173B}
{Brammer} G.~B. {et~al.}, 2009, \apjl, 706, L173

\bibitem[{{Courteau}(1997)}]{1997AJ....114.2402C}
{Courteau} S., 1997, \aj, 114, 2402

\bibitem[{{Courtois} {et~al}\mbox{.}(2012){Courtois}, {Hoffman}, {Tully}, \&
  {Gottl{\"o}ber}}]{2012ApJ...744...43C}
{Courtois} H.~M., {Hoffman} Y., {Tully} R.~B., {Gottl{\"o}ber} S., 2012, \apj,
  744, 43

\bibitem[{{Courtois} \& {Tully}(2012)}]{2012ApJ...749..174C}
{Courtois} H.~M., {Tully} R.~B., 2012, \apj, 749, 174

\bibitem[{{Dale} {et~al}\mbox{.}(2009){Dale}, {Cohen}, {Johnson}, {Schuster},
  {Calzetti}, {Engelbracht}, {Gil de Paz}, {Kennicutt}, {Lee}, {Begum},
  {Block}, {Dalcanton}, {Funes}, {Gordon}, {Johnson}, {Marble}, {Sakai},
  {Skillman}, {van Zee}, {Walter}, {Weisz}, {Williams}, {Wu}, \&
  {Wu}}]{2009ApJ...703..517D}
{Dale} D.~A. {et~al.}, 2009, \apj, 703, 517

\bibitem[{{De Jong}(1996)}]{1996A&AS..118..557D}
{De Jong} R.~S., 1996, \aaps, 118, 557

\bibitem[{{de Jong}(1996)}]{1996A&A...313...45D}
{de Jong} R.~S., 1996, \aap, 313, 45

\bibitem[{{Draine} \& {Lee}(1984)}]{1984ApJ...285...89D}
{Draine} B.~T., {Lee} H.~M., 1984, \apj, 285, 89

\bibitem[{{Fazio} {et~al}\mbox{.}(2004){Fazio}, {Hora}, {Allen}, {Ashby},
  {Barmby}, {Deutsch}, {Huang}, {Kleiner}, {Marengo}, {Megeath}, {Melnick},
  {Pahre}, {Patten}, {Polizotti}, {Smith}, {Taylor}, {Wang}, {Willner},
  {Hoffmann}, {Pipher}, {Forrest}, {McMurty}, {McCreight}, {McKelvey},
  {McMurray}, {Koch}, {Moseley}, {Arendt}, {Mentzell}, {Marx}, {Losch},
  {Mayman}, {Eichhorn}, {Krebs}, {Jhabvala}, {Gezari}, {Fixsen}, {Flores},
  {Shakoorzadeh}, {Jungo}, {Hakun}, {Workman}, {Karpati}, {Kichak}, {Whitley},
  {Mann}, {Tollestrup}, {Eisenhardt}, {Stern}, {Gorjian}, {Bhattacharya},
  {Carey}, {Nelson}, {Glaccum}, {Lacy}, {Lowrance}, {Laine}, {Reach},
  {Stauffer}, {Surace}, {Wilson}, {Wright}, {Hoffman}, {Domingo}, \&
  {Cohen}}]{2004ApJS..154...10F}
{Fazio} G.~G. {et~al.}, 2004, \apjs, 154, 10

\bibitem[{{Freedman} {et~al}\mbox{.}(2001){Freedman}, {Madore}, {Gibson},
  {Ferrarese}, {Kelson}, {Sakai}, {Mould}, {Kennicutt}, {Ford}, {Graham},
  {Huchra}, {Hughes}, {Illingworth}, {Macri}, \&
  {Stetson}}]{2001ApJ...553...47F}
{Freedman} W.~L. {et~al.}, 2001, \apj, 553, 47

\bibitem[{{Freedman} {et~al}\mbox{.}(2011){Freedman}, {Madore}, {Scowcroft},
  {Monson}, {Persson}, {Seibert}, {Rigby}, {Sturch}, \&
  {Stetson}}]{2011AJ....142..192F}
{Freedman} W.~L. {et~al.}, 2011, \aj, 142, 192

\bibitem[{{Freeman}(1970)}]{1970ApJ...160..811F}
{Freeman} K.~C., 1970, \apj, 160, 811

\bibitem[{{Huizinga}(1994)}]{1994essg.book.....H}
{Huizinga} J.~E., 1994, {Extinction studies of spiral galaxies}

\bibitem[{{Lee} {et~al}\mbox{.}(1993){Lee}, {Freedman}, \&
  {Madore}}]{1993ApJ...417..553L}
{Lee} M.~G., {Freedman} W.~L., {Madore} B.~F., 1993, \apj, 417, 553

\bibitem[{{Makarov} {et~al}\mbox{.}(2006){Makarov}, {Makarova}, {Rizzi},
  {Tully}, {Dolphin}, {Sakai}, \& {Shaya}}]{2006AJ....132.2729M}
{Makarov} D., {Makarova} L., {Rizzi} L., {Tully} R.~B., {Dolphin} A.~E.,
  {Sakai} S., {Shaya} E.~J., 2006, \aj, 132, 2729

\bibitem[{{Mart{\'{\i}}nez} {et~al}\mbox{.}(2006){Mart{\'{\i}}nez}, {O'Mill},
  \& {Lambas}}]{2006MNRAS.372..253M}
{Mart{\'{\i}}nez} H.~J., {O'Mill} A.~L., {Lambas} D.~G., 2006, \mnras, 372, 253

\bibitem[{{McDonald} {et~al}\mbox{.}(2009{\natexlab{a}}){McDonald}, {Courteau},
  \& {Tully}}]{2009MNRAS.393..628M}
{McDonald} M., {Courteau} S., {Tully} R.~B., 2009{\natexlab{a}}, \mnras, 393,
  628

\bibitem[{{McDonald} {et~al}\mbox{.}(2009{\natexlab{b}}){McDonald}, {Courteau},
  \& {Tully}}]{2009MNRAS.394.2022M}
{McDonald} M., {Courteau} S., {Tully} R.~B., 2009{\natexlab{b}}, \mnras, 394,
  2022

\bibitem[{{McGaugh}(1996)}]{1996MNRAS.280..337M}
{McGaugh} S.~S., 1996, \mnras, 280, 337

\bibitem[{{McGaugh} {et~al}\mbox{.}(1995){McGaugh}, {Bothun}, \&
  {Schombert}}]{1995AJ....110..573M}
{McGaugh} S.~S., {Bothun} G.~D., {Schombert} J.~M., 1995, \aj, 110, 573

\bibitem[{{Meidt} {et~al}\mbox{.}(2012){Meidt}, {Schinnerer}, {Knapen},
  {Bosma}, {Athanassoula}, {Sheth}, {Buta}, {Zaritsky}, {Laurikainen},
  {Elmegreen}, {Elmegreen}, {Gadotti}, {Salo}, {Regan}, {Ho}, {Madore}, {Hinz},
  {Skibba}, {Gil de Paz}, {Mu{\~n}oz-Mateos}, {Men{\'e}ndez-Delmestre},
  {Seibert}, {Kim}, {Mizusawa}, {Laine}, \&
  {Comer{\'o}n}}]{2012ApJ...744...17M}
{Meidt} S.~E. {et~al.}, 2012, \apj, 744, 17

\bibitem[{{Mestel}(1963)}]{1963MNRAS.126..553M}
{Mestel} L., 1963, \mnras, 126, 553

\bibitem[{{Mosenkov} {et~al}\mbox{.}(2010){Mosenkov}, {Sotnikova}, \&
  {Reshetnikov}}]{2010MNRAS.401..559M}
{Mosenkov} A.~V., {Sotnikova} N.~Y., {Reshetnikov} V.~P., 2010, \mnras, 401,
  559

\bibitem[{{Paturel} {et~al}\mbox{.}(2003){Paturel}, {Petit}, {Prugniel},
  {Theureau}, {Rousseau}, {Brouty}, {Dubois}, \&
  {Cambr{\'e}sy}}]{2003A&A...412...45P}
{Paturel} G., {Petit} C., {Prugniel} P., {Theureau} G., {Rousseau} J., {Brouty}
  M., {Dubois} P., {Cambr{\'e}sy} L., 2003, \aap, 412, 45

\bibitem[{{Reach} {et~al}\mbox{.}(2005){Reach}, {Megeath}, {Cohen}, {Hora},
  {Carey}, {Surace}, {Willner}, {Barmby}, {Wilson}, {Glaccum}, {Lowrance},
  {Marengo}, \& {Fazio}}]{2005PASP..117..978R}
{Reach} W.~T. {et~al.}, 2005, \pasp, 117, 978

\bibitem[{{Rizzi} {et~al}\mbox{.}(2007){Rizzi}, {Tully}, {Makarov}, {Makarova},
  {Dolphin}, {Sakai}, \& {Shaya}}]{2007ApJ...661..815R}
{Rizzi} L., {Tully} R.~B., {Makarov} D., {Makarova} L., {Dolphin} A.~E.,
  {Sakai} S., {Shaya} E.~J., 2007, \apj, 661, 815

\bibitem[{{Schombert}(2007)}]{2007astro.ph..3646S}
{Schombert} J., 2007, ArXiv Astrophysics e-prints

\bibitem[{{Schombert} \& {Smith}(2012)}]{2012PASA...29..174S}
{Schombert} J., {Smith} A., 2012, PASA, 29, 174

\bibitem[{{Sersic}(1959)}]{1959Obs....79...54S}
{Sersic} J.~L., 1959, The Observatory, 79, 54

\bibitem[{{Sheth} {et~al}\mbox{.}(2010){Sheth}, {Regan}, {Hinz}, {Gil de Paz},
  {Men{\'e}ndez-Delmestre}, {Mu{\~n}oz-Mateos}, {Seibert}, {Kim},
  {Laurikainen}, {Salo}, {Gadotti}, {Laine}, {Mizusawa}, {Armus},
  {Athanassoula}, {Bosma}, {Buta}, {Capak}, {Jarrett}, {Elmegreen},
  {Elmegreen}, {Knapen}, {Koda}, {Helou}, {Ho}, {Madore}, {Masters},
  {Mobasher}, {Ogle}, {Peng}, {Schinnerer}, {Surace}, {Zaritsky},
  {Comer{\'o}n}, {de Swardt}, {Meidt}, {Kasliwal}, \&
  {Aravena}}]{2010PASP..122.1397S}
{Sheth} K. {et~al.}, 2010, \pasp, 122, 1397

\bibitem[{{Sorce} {et~al}\mbox{.}(2012{\natexlab{a}}){Sorce}, {Courtois}, \&
  {Tully}}]{2012AJ....144..133S}
{Sorce} J.~G., {Courtois} H.~M., {Tully} R.~B., 2012{\natexlab{a}}, \aj, 144,
  133

\bibitem[{{Sorce} {et~al}\mbox{.}(2013){Sorce}, {Courtois}, {Tully}, {Seibert},
  {Scowcroft}, {Freedman}, {Madore}, {Persson}, {Monson}, \&
  {Rigby}}]{2013ApJ...765...94S}
{Sorce} J.~G. {et~al.}, 2013, \apj, 765, 94

\bibitem[{{Sorce} {et~al}\mbox{.}(2012{\natexlab{b}}){Sorce}, {Tully}, \&
  {Courtois}}]{2012ApJ...758L..12S}
{Sorce} J.~G., {Tully} R.~B., {Courtois} H.~M., 2012{\natexlab{b}}, \apjl, 758,
  L12

\bibitem[{{Stoughton} {et~al}\mbox{.}(2002){Stoughton}, {Lupton}, {Bernardi},
  {Blanton}, {Burles}, {Castander}, {Connolly}, {Eisenstein}, {Frieman},
  {Hennessy}, {Hindsley}, {Ivezi{\'c}}, {Kent}, {Kunszt}, {Lee}, {Meiksin},
  {Munn}, {Newberg}, {Nichol}, {Nicinski}, {Pier}, {Richards}, {Richmond},
  {Schlegel}, {Smith}, {Strauss}, {SubbaRao}, {Szalay}, {Thakar}, {Tucker},
  {Vanden Berk}, {Yanny}, {Adelman}, {Anderson}, {Anderson}, {Annis},
  {Bahcall}, {Bakken}, {Bartelmann}, {Bastian}, {Bauer}, {Berman},
  {B{\"o}hringer}, {Boroski}, {Bracker}, {Briegel}, {Briggs}, {Brinkmann},
  {Brunner}, {Carey}, {Carr}, {Chen}, {Christian}, {Colestock}, {Crocker},
  {Csabai}, {Czarapata}, {Dalcanton}, {Davidsen}, {Davis}, {Dehnen},
  {Dodelson}, {Doi}, {Dombeck}, {Donahue}, {Ellman}, {Elms}, {Evans}, {Eyer},
  {Fan}, {Federwitz}, {Friedman}, {Fukugita}, {Gal}, {Gillespie}, {Glazebrook},
  {Gray}, {Grebel}, {Greenawalt}, {Greene}, {Gunn}, {de Haas}, {Haiman},
  {Haldeman}, {Hall}, {Hamabe}, {Hansen}, {Harris}, {Harris}, {Harvanek},
  {Hawley}, {Hayes}, {Heckman}, {Helmi}, {Henden}, {Hogan}, {Hogg}, {Holmgren},
  {Holtzman}, {Huang}, {Hull}, {Ichikawa}, {Ichikawa}, {Johnston}, {Kauffmann},
  {Kim}, {Kimball}, {Kinney}, {Klaene}, {Kleinman}, {Klypin}, {Knapp},
  {Korienek}, {Krolik}, {Kron}, {Krzesi{\'n}ski}, {Lamb}, {Leger},
  {Limmongkol}, {Lindenmeyer}, {Long}, {Loomis}, {Loveday}, {MacKinnon},
  {Mannery}, {Mantsch}, {Margon}, {McGehee}, {McKay}, {McLean}, {Menou},
  {Merelli}, {Mo}, {Monet}, {Nakamura}, {Narayanan}, {Nash}, {Neilsen},
  {Newman}, {Nitta}, {Odenkirchen}, {Okada}, {Okamura}, {Ostriker}, {Owen},
  {Pauls}, {Peoples}, {Peterson}, {Petravick}, {Pope}, {Pordes}, {Postman},
  {Prosapio}, {Quinn}, {Rechenmacher}, {Rivetta}, {Rix}, {Rockosi}, {Rosner},
  {Ruthmansdorfer}, {Sandford}, {Schneider}, {Scranton}, {Sekiguchi}, {Sergey},
  {Sheth}, {Shimasaku}, {Smee}, {Snedden}, {Stebbins}, {Stubbs}, {Szapudi},
  {Szkody}, {Szokoly}, {Tabachnik}, {Tsvetanov}, {Uomoto}, {Vogeley}, {Voges},
  {Waddell}, {Walterbos}, {Wang}, {Watanabe}, {Weinberg}, {White}, {White},
  {Wilhite}, {Wolfe}, {Yasuda}, {York}, {Zehavi}, \&
  {Zheng}}]{2002AJ....123..485S}
{Stoughton} C. {et~al.}, 2002, \aj, 123, 485

\bibitem[{{Swaters} {et~al}\mbox{.}(2010){Swaters}, {Sanders}, \&
  {McGaugh}}]{2010ApJ...718..380S}
{Swaters} R.~A., {Sanders} R.~H., {McGaugh} S.~S., 2010, \apj, 718, 380

\bibitem[{{Thompson}(2003)}]{2003Ap&SS.284..353T}
{Thompson} R.~I., 2003, \apss, 284, 353

\bibitem[{{Tully} \& {Courtois}(2012)}]{2012ApJ...749...78T}
{Tully} R.~B., {Courtois} H.~M., 2012, \apj, 749, 78

\bibitem[{{Tully} \& {Fisher}(1977)}]{1977A&A....54..661T}
{Tully} R.~B., {Fisher} J.~R., 1977, \aap, 54, 661

\bibitem[{{Tully} {et~al}\mbox{.}(2009){Tully}, {Rizzi}, {Shaya}, {Courtois},
  {Makarov}, \& {Jacobs}}]{2009AJ....138..323T}
{Tully} R.~B., {Rizzi} L., {Shaya} E.~J., {Courtois} H.~M., {Makarov} D.~I.,
  {Jacobs} B.~A., 2009, \aj, 138, 323

\bibitem[{{Tully} \& {Verheijen}(1997)}]{1997ApJ...484..145T}
{Tully} R.~B., {Verheijen} M.~A.~W., 1997, \apj, 484, 145

\bibitem[{{Werner} {et~al}\mbox{.}(2004){Werner}, {Roellig}, {Low}, {Rieke},
  {Rieke}, {Hoffmann}, {Young}, {Houck}, {Brandl}, {Fazio}, {Hora}, {Gehrz},
  {Helou}, {Soifer}, {Stauffer}, {Keene}, {Eisenhardt}, {Gallagher}, {Gautier},
  {Irace}, {Lawrence}, {Simmons}, {Van Cleve}, {Jura}, {Wright}, \&
  {Cruikshank}}]{2004ApJS..154....1W}
{Werner} M.~W. {et~al.}, 2004, \apjs, 154, 1

\bibitem[{{Wetzel} {et~al}\mbox{.}(2012){Wetzel}, {Tinker}, \&
  {Conroy}}]{2012MNRAS.424..232W}
{Wetzel} A.~R., {Tinker} J.~L., {Conroy} C., 2012, \mnras, 424, 232

\bibitem[{{Whitaker} {et~al}\mbox{.}(2011){Whitaker}, {Labb{\'e}}, {van
  Dokkum}, {Brammer}, {Kriek}, {Marchesini}, {Quadri}, {Franx}, {Muzzin},
  {Williams}, {Bezanson}, {Illingworth}, {Lee}, {Lundgren}, {Nelson},
  {Rudnick}, {Tal}, \& {Wake}}]{2011ApJ...735...86W}
{Whitaker} K.~E. {et~al.}, 2011, \apj, 735, 86

\bibitem[{{Zhong} {et~al}\mbox{.}(2008){Zhong}, {Liang}, {Liu}, {Hammer}, {Hu},
  {Chen}, {Deng}, \& {Zhang}}]{2008MNRAS.391..986Z}
{Zhong} G.~H., {Liang} Y.~C., {Liu} F.~S., {Hammer} F., {Hu} J.~Y., {Chen}
  X.~Y., {Deng} L.~C., {Zhang} B., 2008, \mnras, 391, 986

\end{thebibliography}

\end{document}